\documentclass[12pt]{article}
\usepackage{standard}
\usepackage{psfig}
\newcommand{\Struta}{\rule{0in}{3ex}}

\newenvironment{Eqnarray}%
    {\arraycolsep 0.14em\begin{eqnarray}}{\end{eqnarray}}
\newenvironment{Eqnarray*}%
    {\arraycolsep 0.14em\begin{eqnarray*}}{\end{eqnarray*}}

\begin{document}
\thispagestyle{empty}
\begin{flushright}
        MZ-TH/98-47\\
        hep-ph/9811482\\
        November 1998\\
\end{flushright}
\vspace{0.5cm}
\begin{center}
{\Large\bf Polarized top decay into polarized W}\\[.3cm]
{\Large\bf $t(\uparrow) \rightarrow W(\uparrow) \!+\! b$ at 
           $ O(\alpha_s) $}\\[0.7cm]
{\large M. Fischer, S.~Groote, J.G.~K"orner and M.C.~Mauser}\\[0.7cm]
        Institut f"ur Physik, Johannes-Gutenberg-Universit"at\\[.2cm]
        Staudinger Weg 7, D--55099 Mainz, Germany\\[0.7cm]
        and\\[0.7cm]
{\large B.~Lampe}\\[0.7cm]
        Sektion Physik\\[0.2cm]
        Ludwig--Maximilians--Universit"at M"unchen\\[.2cm]
        Theresienstra"se 37, D--80333 M"unchen\\[0.2cm] and \\[0.2cm] 
        MPI f"ur Physik, Werner--Heisenberg--Institut\\[.2cm]
        F"ohringer Ring 6, D--80805 M"unchen
\end{center}
\vspace{0.7cm}


\begin{abstract}\noindent
  We consider the decay of a polarized top quark into a polarized $ W $-boson
  plus a bottom quark, followed by the decay of the $ W $-boson into a pair of
  leptons or quarks. The polar angle distribution  of the top spin relative
  to the $W$-momentum and the polar angle distribution of the lepton (or quark)
  in the $W$-rest frame is governed by three polarized and three unpolarized
  rate functions which are related to the double density matrix elements of the
  decay $ t \!\rightarrow\! W^+ + b $. We obtain analytical expressions for the
  $ O(\alpha_s) $ radiative corrections to the three polarized and three 
  unpolarized rate functions. We also provide a comprehensive discussion
  of the dependence of the longitudinal, transverse and normal polarization of
  the top quarks produced at $e^+e^-$-colliders on beam polarization parameters.
\end{abstract}

\newpage

  
  It is well-known that quarks produced in $ e^+ e^- $-annihilation possess a 
  high degree of polarization. This holds true for the production of light 
  quarks $ (u,d,s,c,b) $ as well as for heavy top quarks. The top quark is very
  short-lived and therefore retains its full polarization content when it 
  decays. By measuring the momenta and spin orientation of its decay products
  one can then define spin-momentum and spin-spin correlations between the top
  quark spin and its decay products which will allow for detailed studies of the
  top decay mechanism.

  In this paper we study momentum and spin--momentum correlations in the cascade
  decay process $ t \!\rightarrow\! W^+ \!+\! b $ followed by $ W^+ \!\rightarrow\!
  l^+ \!+\! \nu_l $ or $ W \!\rightarrow\! \bar{q} \!+\! q $. The step--one decay
  $ t \!\rightarrow\! W^+ \!+\! b $ is analyzed in the $ t $-rest frame where we
  study the spin-momentum correlation between the spin of the top and the momentum
  of the $ W $. In step two we go to the rest frame of the $ W $ and analyze the
  correlation between the momentum of the lepton (or antiquark) and the initial
  momentum of the $ W $, i.e. we thereby analyze the spin density matrix of the
  $ W $.

  One of the attractions of the envisaged future high energy linear colliders is
  the possibility to tune the polarization of the top quark over a wide range of
  polarization values by making use of the possibility of linear colliders to
  polarize their $e^+$ and $e^-$ beams. Prospective linear $e^+e^-$-colliders
  are planned to have available maximal polarizations of $ P(e^-) \!=\! 80 \!-\!
  90 \%$ and $ P(e^+) \!=\! 60 \!-\! 70 \%$.

  Before we turn to the main subject of the paper, namely the decay of a 
  polarized top quark into a polarized $ W $ and a (massless) bottom quark, we
  want to first demonstrate the capabilities of linear colliders to tune the 
  polarization of the top quarks through the use of beam polarization. To this
  end we write down the functional dependence of the mean longitudinal and
  transverse components of the top quark's polarization on the beam polarization
  parameters. Here the mean refers to polar angle averaging over the polar angle
  between the beam and the top quark. In the Standard Model and at the Born term
  level one has (see e.g. \cite{c1,c2,c3,c4})
  \begin{equation} 
     P^L = \frac{4}{3} \frac{
     [h_1 g_{14} \!+\! h_2 g_{44}] v}{
     [h_1 g_{11} \!+\! h_2 g_{41}] (1 \!+\! \frac{1}{3} v^2) \!+\! 
     [h_1 g_{12} \!+\! h_2 g_{42}] (1 \!-\! v^2)}
  \end{equation} 
  and 
  \begin{equation} 
     P^{\perp} = - \frac{\pi}{2} \frac{m_t}{\sqrt{q^2}} \frac{
      h_2 (g_{11} \!+\! g_{12}) \!+\! h_1 (g_{41} \!+\! g_{42})}{
     [h_1 g_{11} \!+\! h_2 g_{41}] (1 \!+\! \frac{1}{3} v^2) \!+\!
     [h_1 g_{12} \!+\! h_2 g_{42}] (1 \!-\! v^2)}
  \end{equation}
  where $ v \!=\! (1 \!-\! 4 m^2/ q^2)^{1/2} $, $ h_1 \!=\! 1 \!-\! h^- h^+ $,
  $ h_2 \!=\! h^- \!-\! h^+ $ and $ h^- $ and $ h^+ $ are the longitudinal (or
  helicity) polarizations of the $ e^- $ and $ e^+ $ beams, respectively. The
  electroweak model dependence is specified by the electroweak model parameters
  $ g_{ij} $ which, in the Standard Model, read~\cite{c3}
  \begin{Eqnarray}
     g_{11} & = & \frac{4}{9} \!-\! \frac{4}{3} \; v_e v_t \; \mbox{Re}(\chi_Z)
      \!+\! (v_e^2 \!+\! a_e^2)(v_t^2 \!+\! a_t^2) \; |\chi_Z|^2 \\
     g_{12} & = & \frac{4}{9} \!-\! \frac{4}{3} \; v_e v_t \; \mbox{Re}(\chi_Z) 
      \!+\! (v_e^2 \!+\! a_e^2)(v_t^2 \!-\! a_t^2) \; |\chi_Z|^2 \\
     g_{14} & = & \frac{4}{3} \; v_e a_t \; \mbox{Re}(\chi_Z) 
      \!-\! 2 (v_e^2 \!+\! a_e^2) \; v_t a_t \; |\chi_Z|^2 \\
     g_{41} & = & \frac{4}{3} \; a_e v_t \; \mbox{Re}(\chi_Z) 
      \!-\! 2 (v_t^2 \!+\! a_t^2) \; v_e a_e \; |\chi_Z|^2 \\
     g_{42} & = & \frac{4}{3} \; a_e v_t \; \mbox{Re}(\chi_Z)
      \!-\! 2 (v_t^2 \!-\! a_t^2) \; v_e a_e \; |\chi_Z|^2 \\
     g_{44} & = & - \frac{4}{3} \; a_e a_t \; \mbox{Re}(\chi_z)
      \!+\! 4 \; v_e a_e v_t a_t \; |\chi_Z|^2                   
  \end{Eqnarray}
  \begin{equation}
     v_e = \!-\! 1 \!+\! 4 \sin^2 \theta_W, \quad a_e = \!-\! 1, \quad
     v_t = 1 \!-\! \frac{8}{3} \sin^2 \theta_W, \quad a_t = 1
  \end{equation}
  and where the propagator function $ \chi_Z $ reads
  \begin{equation} 
     \chi_Z(s) =  g M_Z^2 \frac{s}{s - M_Z^2 + i M_Z \Gamma_Z}.
  \end{equation}
  $ M_Z $ and $ \Gamma_Z $ are the mass and width of the Z-boson and $ g \!=\!
  G_F (8 \sqrt{2} \pi \alpha)^{-1} \!\approx\! 4.49 \cdot\! 10^{-5} \mbox{
  GeV}^{-2} $. We take $ \sin^2 \theta_W \!=\! 0.23124 $ from \cite{cj}.

  At the Born term level the polarization lies entirely in the plane spanned by
  the beam and the top quark (we are neglecting an imaginary contribution from
  the Breit-Wigner form of the propagator function). In Fig.~1 we plot the two 
  non-vanishing components of the top polarization $P^L$ and $P^{\perp}$ as a
  function of the center of mass energy $ \sqrt{q^2} $ for $ h^+ \!=\! 0 $, and
  $ h^- $ ranges from $ -1 $ to $ +1 $ in increments of $ 0.2 $. The boundary
  curves $h^- \!=\! -1$ and $ h^- \!=\! +1 $ for both the longitudinal and
  transverse polarization can be seen to be independent of the value of $ h^+ $.
  At $ \sqrt{q^2} \!=\! 500 \; \mbox{GeV} $ the longitudinal polarization 
  becomes as large as $ P^L \!=\! -52.2 \% $ and $ P^L \!=\! +65.2 \% $ for
  $ h^- \!=\! -1 $ and $ h^- \!=\! +1$, respectively. The value of $ h^+ $ 
  determines the density of the $ h^- $-lines as the region between the two
  boundary curves is traversed. For $ h^+ \!\approx\! -0.3 $ the different
  $ h^- $-lines become approximately equally spaced. For $ h^+ \!\rightarrow\!
  -1 $ the different $ h^- $-lines migrate towards the upper boundary curve
  $ h^- \!=\! +1 $, whereas they migrate towards the lower boundary curve
  $ h^- \!=\! -1 $ for $ h^+ \!\rightarrow\! +1 $.
 
  The case $ h^+ \!=\! 0 $ depicted in Fig.~1 provides a good demonstration of
  the possibilities of polarization tuning at $ e^+e^- $-colliders. In addition
  it does not require the positron beam to be polarized which is an advantage. 
  The transverse polarization shown in Fig.~1 is generally comparable in size to
  the longitudinal polarization. At $ \sqrt{q^2} \!=\! 500 \; \mbox{GeV} $ the
  transverse polarization is $ P^{\perp} \!=\! -1.8 \% $ and $ P^{\perp} \!=\!
  -2.2 \% $ for $ h^- \!=\! -1 $ and $ h^- \!=\! +1 $, respectively. It shows
  the expected $ 1/\sqrt{q^2} $ fall-off behaviour while the longitudinal
  polarization slightly increases with energy. We want to remind the reader that
  we have calculated the mean polarization of the top after averaging over the
  relative orientation between the beam and the top quark. In fact, the $ \cos
  \theta $-dependence of the polarization can be quite strong in particular for
  small values of the mean polarization \cite{c1,c2,c3,c4}.
  
  The $ O(\alpha_s) $ radiative corrections to the Born term polarization 
  have been calculated in \cite{c3,c4,c5a,c5b}. They are generally quite small.
  Also at $ O(\alpha_s) $ a small polarization contribution normal
  to the beam-event plane comes in through the imaginary part of the one-loop
  contribution \cite{c1,c4}. For the normal polarization $ P^N $ one 
  obtains\footnote{There is a factor $ (v_e^2 + a_e^2)^{-1} $ missing on the
  r.h.s of Eq.(16) in \cite{c1}.} \cite{c1,c4}
  \begin{equation} 
     P^{N} = - \frac{\pi}{6} \frac{m_t}{\sqrt{q^2}} (2 \!-\! v^2) \alpha_s
     \frac{h_2 g_{14} \!+\! h_1 g_{44}}{
     [h_1 g_{11} \!+\! h_2 g_{41}] (1 \!+\! \frac{1}{3} v^2) \!+\!
     [h_1 g_{12} \!+\! h_2 g_{42}] (1 \!-\! v^2)}.
  \end{equation}
  In Fig.~2 we plot the normal component of the top polarization $ P^N $ as a
  function of the center of mass energy again for $ h^+ \!=\! 0 $, where $ h^- $
  ranges from $ h^- \!=\! -1 $ to $ h^+ \!=\! 1 $ in increments of 1. As before
  the boundary curves for $ h^- \!=\! -1 $ and $ h^- \!=\! 1 $ are independent
  of $ h^+ $ as a simple inspection of Eq.~(11) shows. Contrary to e.g. 
  $ P^{L} $ in Eq.~(1) the numerator in Eq.~(11) determining $ P^N $ for the
  two boundary curves is now given by $ \pm (g_{14} \!\pm\! g_{44}) $ and not
  by $ (g_{14} \!\pm\! g_{44}) $. From the values of $ g_{14} $ and $ g_{44} $
  this implies that the boundary curves for $ P^N $ are not as spread apart as
  for $ P^L $. The numerical value of $ P^N $ is generally quite small which,
  in part, is due to the overall factor $ \alpha_s $ multiplying $ P^N $. The
  normal component $ P^N $ shows the expected $ 1/\sqrt{q^2} $ fall-off 
  behaviour.

  We now turn to the main subject of this paper, namely the decay of a 
  polarized top quark into an on-shell polarized $ W^+ $ and a massless
  bottom quark. The polarization of the $ W^+ $ can be probed through the
  angular decay distribution of its decay products, which can be a lepton pair 
  $ (l^+,\nu_l) $ or a light quark-antiquark pair. In Fig.2 we introduce the
  two polar angles $ \theta_P $ and $ \theta $. They are defined in the two
  respective rest frames of the two-step cascade process $ t \rightarrow W^+
  + b $ followed by $ W^+ \!\rightarrow\! l^+ \!+\! \nu_l $ or $ W^+ 
  \!\rightarrow\! \bar{q} \!+\! q $. The double polar angle decay distribution
  is given by (see e.g. \cite{c6})
  \begin{Eqnarray}
    \frac{1}{\Gamma_0} \frac{d \Gamma}{d \cos \theta_P d \cos \theta} & = &
    \frac{1}{2} \Big( \hat{\Gamma}_U + \hat{\Gamma}_U^P \; P \cos \theta_P \Big)
    \frac{3}{8} \Big(1 \!+\! \cos^2 \theta \Big) \nonumber \\ & + & 
    \frac{1}{2} \Big( \hat{\Gamma}_L + \hat{\Gamma}_L^P \; P \cos \theta_P \Big)
    \frac{3}{4} \sin^2 \theta \\ & + &
    \frac{1}{2} \Big( \hat{\Gamma}_F + \hat{\Gamma}_F^P \; P \cos \theta_P \Big)
    \frac{3}{4} \cos \theta. \nonumber
  \end{Eqnarray}

  \noindent As a reference rate we take the total Born term rate $ \Gamma_0 \!=\!
  \Gamma_{U+L}(\mbox{Born}) $ given by ($ x \!=\! m_W/m_t $)
  \begin{equation}
    \Gamma_0 = \frac{G_F m_W^2  m_t}{4 \sqrt{2} \pi} |V_{tb}|^2
    \frac{(1 \!-\! x^2)^2 (1 \!+\! 2x^2)}{x^2}.
  \end{equation}
  We have defined scaled rate functions $ \hat{\Gamma}_i $ and $ \hat{\Gamma}_i^P
  $ according to $ \hat{\Gamma}_i \!=\! \Gamma_i/\Gamma_0 $ and $ \hat{\Gamma}_i^P
  \!=\! \Gamma_i^P/\Gamma_0 $. The subscripts in the unpolarized rate functions
  $ \Gamma_i $ and the polarized rate functions $ \Gamma_i^P $ ($ i \!=\! U,L,F $)
  refer to three measurable polarization states of the $ W^+ $ ($U$: unpolarized
  transverse; $L$: longitudinal; $F$: forward-backward asymmetric). For example, for
  the unpolarized rate functions $ \Gamma_i $ ($ i \!=\! U,L,F $) the three
  components are determined by the three linear combinations $ \Gamma_U \!=\!
  \Gamma_{++} \!+ \Gamma_{--} $, $ \Gamma_L \!=\! \Gamma_{\; 00 \;} $ and $
  \Gamma_F \!=\! \Gamma_{++} \!-\! \Gamma_{--} $ of the diagonal density matrix
  elements  $ \Gamma_{\lambda_W, \lambda_W} $ of the $ W $ ($ \lambda_W \!=\!
  \lambda'_W \!=\! 0, \pm 1 $), and similarly for the polarized rate functions.
  The total rate is obviously given by $ \Gamma \!=\! \Gamma_U + \Gamma_L
  \!:=\! \Gamma_{U+L} $ as is evident from Eq.~(12) after $ \cos \theta $ and 
  $ \cos \theta_P $ integration.
  For the unpolarized rate functions one sums over the two diagonal density
  matrix elements of the top quark whereas one takes the difference of the two
  for the polarized rate functions. In practise this is achieved by replacing
  the spin sum $ (\slash{p}_t \!+\! m_t) $ by  $ (\slash{p}_t \!+\! m_t) \gamma_5
  \slash{s}_t $ when calculating the polarized rate functions, where $ s^{\mu}_t
  \!=\! (0,0,0,1) $ in the rest frame of the top quark.

  Let us first list the Born term contributions to the rate functions. They
  are given by
  \begin{Eqnarray}
    \hat{\Gamma}_U (\mbox{Born}) & = & \frac{2 x^2}{1 \!+\! 2 x^2} \quad\quad
    \hat{\Gamma}_U^P (\mbox{Born}) = \frac{- 2 x^2}{1 \!+\! 2 x^2} \nonumber \\
    \hat{\Gamma}_L (\mbox{Born}) & = & \frac{1}{1 \!+\! 2 x^2} \quad\quad
    \hat{\Gamma}_L^P (\mbox{Born}) = \frac{1}{1 \!+\! 2 x^2} \\
    \hat{\Gamma}_F (\mbox{Born}) & = & \frac{- 2 x^2}{1 \!+\! 2 x^2} \quad\quad
    \hat{\Gamma}_F^P (\mbox{Born}) = \frac{2 x^2}{1 \!+\! 2 x^2} \nonumber
  \end{Eqnarray}

  \noindent The reason that one has $\hat{\Gamma}_U \!=\! -\hat{\Gamma}_F \!=\!
  -\hat{\Gamma}_U^P \!=\! \hat{\Gamma}_F^P $ and $\hat{\Gamma}_L \!=\! \hat{
  \Gamma}_L^P $ at the Born term level is that, with a massless $ b $-quark and
  a $ (V \!-\! A) $ interaction, only the two helicity configurations $ (
  \lambda_t \!=\! -1/2;\lambda_W \!=\! -1) $ and $ (\lambda_t \!=\! 1/2;
  \lambda_W \!=\! 0) $ can be realized. For antitop decay one has the two
  helicity configurations ($ \lambda_{\bar{t}} \!=\! 1/2; \lambda_W \!=\! 1)$
  and ($ \lambda_{\bar{t}} \!=\! - 1/2; \lambda_W \!=\! 0 $) such that $
  \hat{\Gamma}_U^P \!\rightarrow\! - \hat{\Gamma}_U^P $, $ \hat{\Gamma}_L^P
  \!\rightarrow \!-\hat{\Gamma}_L^P $, $ \hat{\Gamma}_F \!\rightarrow\!
  -\hat{\Gamma}_F $, while the other rate functions remain unchanged.
    
  The $ O(\alpha_s) $ radiative corrections to the Born term contributions are
  determined by the sum of the one-loop contributions and the tree-graph
  contributions. Each of the two classes of contributions are infrared (IR) and
  mass (M) singular but the singularities cancel in the sum. As a regularization
  procedure for the regularization of the IR/M singularities we have chosen to
  endow the gluon and the bottom quark with (small) masses $ m_g $ and $ m_b $.
  We have given preference to mass regularization over dimensional regularization
  because we wanted to circumvent the notorious $ \gamma_5 $ problem in
  dimensional regularization which comes in for the polarization observables
  calculated in this paper. The IR/M singularities show up as powers of
  logarithms (up to second order) of the mass regulators. The logarithmic
  contributions cancel in the sum of the one-loop and tree-graph contributions
  and one remains with a finite result. All of this is by now standard procedure
  and we shall not dwell on the subject of IR/M regularization any further. The
  details of the calculation will be presented in a forthcoming publication. We
  reiterate that our results are given for $ m_b \!=\! 0 $. We expect that the
  $ m_b \!\neq\! 0 $ corrections are of the same size as in the Born term 
  contributions where they amount to $ 0.17 \% $ in the total rate.

  Including the Born term contributions the full $O(\alpha_s)$ results are
  given by   


  \begin{Eqnarray}
    \hat{\Gamma}_U & = & \frac{2 x^2}{1 \!+\! 2 x^2} \!+\! \frac{\alpha_s}
    {2 \pi} C_F \frac{x^2}{(1 \!-\! x^2)^2 (1 \!+\! 2 x^2)} \Bigg\{ 
    (x^2 \!-\! 1) (19 \!+\! x^2) + \frac{2}{3} \pi^2 (5 \!+\! 5 x^2 \!-\!
    2 x^4) \nonumber \\ & - & \hspace{-1.8pt}
    2 \frac{(1 \!-\! x^2)^2 (1 \!+\! 2 x^2)}{x^2} \ln (1 \!-\! x^2) \!-\!
    4 (5 \!+\! 7 x^2 \!-\! 2 x^4) \ln (x) \!-\! 2 \frac{(1 \!-\! x)^2
    (5 \!+\! 7 x^2 \!+\! 4 x^3)}{x} \nonumber \\ & \times &
    \ln(x) \ln (1 \!-\! x) \!+\! 2 \frac{(1 \!+\! x)^2 (5 \!+\! 7 x^2 \!-\!
    4 x^3)}{x} \ln(x) \ln(1 \!+\! x) \!-\! (5 \!+\! 4 x \!+\! 15 x^2 \!+\!
    8 x^3) \nonumber \\ & \times & 
    2 \frac{(1 \!-\! x)^2}{x} \, \mbox{Li}_2\,(x) + 2 \frac{(1 \!+\! x)^2
    (5 \!-\! 4 x \!+\! 15 x^2 \!-\! 8 x^3)}{x} \mbox{Li}_2\,(-x) \Bigg\}\\[0.6cm]
    \hat{\Gamma}_{L} & = & \frac{1}{1 \!+\! 2 x^2} \!+\! \frac{\alpha_s}
    {2 \pi} C_F \frac{x^2}{(1 \!-\! x^2)^2 (1 \!+\! 2 x^2)} \Bigg\{ \frac{
    (1 \!-\! x^2 )(5 \!+\! 47 x^2 \!-\! 4 x^4)}{2 x^2} - \frac{2}{3} \pi^2 
    \times \nonumber \\ & \times &
    \frac{(1 \!+\! 5 x^2 \!+\! 2 x^4)}{x^2} \!-\!
    \frac{3 (1 \!-\! x^2)^2}{x^2} \ln (1 \!-\! x^2) \!+\! 16 (1 \!+\! 2 x^2)
    \ln(x) \!-\! (2 \!-\! x \!+\! 6 x^2 \!+\! x^3) 
    \nonumber \\ & \times &
    2 \frac{(1 \!-\! x)^2}{x^2} \ln (1 \!-\! x) \ln(x) - 
    2 \frac{(1 \!+\! x)^2 (2 \!+\! x \!+\! 6 x^2 \!-\! x^3)}{x^2} \ln(x)
    \ln (1 \!+\! x) \nonumber \\ & - &
    2 \frac{(1 \!-\! x)^2 (4 \!+\! 3 x \!+\! 8 x^2
    \!+\! x^3)}{x^2} \mbox{Li}_2(x) \!-\!
    2 \frac{(1 \!+\! x)^2 (4 \!-\! 3 x \!+\! 8 x^2 \!-\! x^3)}{x^2} \,
    \mbox{Li}_2 (-x) \Bigg\} \\[0.6cm]
    \hat{\Gamma}_{F} & = & \frac{-2 x^2}{1 \!+\! 2 x^2} \!+\! \frac{\alpha_s}
    {2 \pi} C_F \frac{x^2}{(1 \!-\! x^2)^2 (1 \!+\! 2 x^2)} \Bigg\{
    2 (1 \!-\! x)^2 (4 x \!-\! 3) \,+\, \frac{2}{3} \pi^2 (2 \!+\! x^2) 
    \nonumber \\ & + & 
    2 \frac{(1 \!-\! x^2)^2 (1 \!+\! 2 x^2)}{x^2} \ln(1 \!-\! x) \,+\,
    2 \frac{(1 \!-\! x^2) (1 \!-\! 9 x^2 \!+\! 2 x^4)}{x^2} \ln (1 \!+\! x)
    \nonumber \\ & + & 
    8 (1 \!-\! x^2)^2 \, \mbox{Li}_2(x) \,+\, 8 (1 \!+\! 3 x^2 \!-\! x^4) \,
    \mbox{Li}_2(-x) \Bigg\} \\[0.6cm]
    \hat{\Gamma}_{U}^{P} & = & \frac{- 2 x^2}{1 \!+\! 2 x^2} \!+\! 
    \frac{\alpha_s}{2 \pi} C_F \frac{x^2}{(1 \!-\! x^2)^2 (1 \!+\! 2 x^2)}
    \Bigg\{ \frac{(1 \!-\! x)^2(\!-\! 12 \!+\! 55 x \!-\! 6 x^2 \!+\! x^3)}{x}
    \!-\! \frac{10}{3} \pi^2 \times \nonumber \\ & \times &
    (2 \!+\! x^2) \!+\! 2 \frac{(1 \!-\! x^2)^2 (1 \!+\! 2 x^2)}{x^2}
    \ln (1 \!-\! x) \!+\! 2 \frac{(1 \!-\! x^2)(7 \!+\! 21 x^2 \!+\! 2 x^4)}{x^2}
    \ln (1 \!+\! x) \nonumber \\ & + &
    8 (1 \!-\! x^2)^2 \, \mbox{Li}_2(x) \!-\! 8 ( 11 \!+\! 3 x^2 \!+\! x^4 ) \,
    \mbox{Li}_2 (-x) \Bigg\} \\[0.6cm]
    \hat{\Gamma}_{L}^{P} & = & \frac{1}{1 \!+\! 2 x^2} \!+\! \frac{\alpha_s}
    {2 \pi} C_F \frac{x^2}{(1 \!-\! x^2)^2 (1 \!+\! 2 x^2)} \Bigg\{ -
    (15 \!-\! 22 x \!+\! 105 x^2 \!-\! 24 x^3 \!+\! 4 x^4) \times
    \nonumber \\ & \times &
    \frac{(1 \!-\! x)^2}{2 x^2}  \!+\! \frac{1}{3} \pi^2 \frac{(1 \!+\! 24 x^2
    \!+\! 10 x^4)}{x^2} \!-\! 3 \frac{(1 \!-\! x^2)^2}{x^2} \ln(1 \!-\! x) 
    \!-\! \frac{(1 \!-\! x^2) (17 \!+\! 53 x^2)}{x^2} 
    \times \nonumber \\ & \times &
    \ln (1 \!+\! x) \!-\! 4 \frac{(1 \!-\! x^2)^2}{x^2} \, \mbox{Li}_2 (x)
    \!+\! 4 \frac{(2 \!+\! 22 x^2 \!+\! 11 x^4)}{x^2} \, \mbox{Li}_2 (-x)
    \Bigg\} \\[0.6cm]
    \hat{\Gamma}_{F}^{P} & = & \frac{2 x^2}{1 \!+\! 2 x^2} \!+\! \frac{\alpha_s}
    {2 \pi} C_F \frac{x^2}{(1 \!-\! x^2)^2 (1 \!+\! 2 x^2)} \Bigg\{
    2 (1 \!-\! x^2) (4 \!+\! x^2) \!-\! \frac{2}{3} \pi^2  (1 \!+\! x^2 \!+\!
    2 x^4) \\ & - &
    2 \frac{(1 \!-\! x^2)^2 (1 \!+\! 2 x^2)}{x^2} \ln (1 \!-\! x^2) - 
    4 (2 \!-\! 5 x^2 \!-\! 2 x^4) \ln(x) - \ln (x) \ln (1 \!-\! x) 
    \times \nonumber \\ & \times &
    4 \frac{(1 \!-\! x)^2 (1 \!+\! 3 x \!+\! 2 x^2 \!+\! 2 x^3)}{x} \!+\!
    4 \frac{(1 \!+\! x)^2 (1 \!-\! 3 x \!+\! 2 x^2 \!-\! 2 x^3)}{x} \ln (x) 
    \ln (1 \!+\! x) \nonumber \\ & - & 
    4 \frac{(1 \!-\! x )^2 (1 \!+\! 5 x \!+\! 6 x^2 \!+\!
    4 x^3)}{x} \mbox{Li}_2 (x) \!+\! 4 \frac{(1 \!+\! x)^2 (1 \!-\! 5 x \!+\!
    6 x^2 \!-\! 4 x^3)}{x} \mbox{Li}_2 (-x) \! \Bigg\} \hspace{0.5cm} \nonumber
  \end{Eqnarray}

  Analytical results have been given before on the total rate $ \Gamma_{U+L} $
  ~\cite{c7,c8,c9}\footnote{The total rate also follows from the work of 
  \cite{ca}} and on the ratio $ \Gamma_U/\Gamma_L $~\cite{cb}. We find full
  agreement with these results. Our analytical results on $ \Gamma_F $ and the
  polarized rate functions $ \Gamma^P_i $ are new. We have also compared our
  results with the numerical results on the remaining rate functions given in
  \cite{cc,cd}. Again we find agreement.

  We now discuss our numerical results. The strong coupling constant has been
  evolved from the $ M_Z $-scale ($ \alpha_s(M_Z) \!=\! 0.1175 $) to the top
  quark mass scale. Thus we take $ \alpha_s(m_t \!=\! 175 \mbox{~GeV}) \!=\!
  0.10702 $ with $ m_W \!=\! 80.22 \mbox{~GeV} $. The numerical values for the
  Born term and the $ O(\alpha_s) $ corrections are given by
  \begin{center}\hfill
  \begin{tabular}{l@{=}rcl@{=}r}
    $ \hat{\Gamma}_U   $ & $  0.296 (1-0.062) $ & \hspace{1cm} &
    $ \hat{\Gamma}_U^P $ & $ -0.296 (1-0.069) $ \\ \Struta
    $ \hat{\Gamma}_L   $ & $  0.704 (1-0.095) $ & \hspace{1cm} &
    $ \hat{\Gamma}_L^P $ & $  0.704 (1-0.096) $ \\ \Struta
    $ \hat{\Gamma}_F   $ & $ -0.296 (1-0.069) $ & \hspace{1cm} &
    $ \hat{\Gamma}_F^P $ & $  0.296 (1-0.064) $.
  \end{tabular}
  \vspace{-1.0truecm} \vspace{-12pt}
  \stepcounter{equation}
  \hfill$(\arabic{equation})$\break
  \vspace{+1.2truecm}
  \end{center}
 
  The radiative corrections to the unpolarized and polarized rate functions
  are sizeable and range from $ 6.2 \% $ to $ 9.6 \% $. However, the radiative
  corrections all go in the same direction. This is an indication that the main
  contributions to the radiative corrections come from the tree-diagrams from
  phase-space regions close to the IR/M singularities. Similar observations have
  been made for the radiative corrections to polarization-type structure
  functions in other production or decay processes. When forming ratios of the
  rate functions, as is appropriate for the definition of polarization-type
  observables, the size of the radiative corrections to the  polarization-type 
  observables are much reduced. For example, the $ O(\alpha_s) $ radiative
  corrections to the ratio $ \Gamma_U/\Gamma_L $ and $ \Gamma_F/\Gamma_{U+L} $
  are reduced to $ 3.65 \% $ and $ 1.77 \% $, respectively.

  In conclusion, we have obtained analytical results on the $ O(\alpha_s) $ 
  corrections to the six unpolarized and polarized rate functions that describe
  the double polar angle distribution of the decay of a polarized top quark 
  into a $ W $ and a massless $ b $-quark followed by the decay of the $ W $
  into a lepton pair or a quark pair. For antitop decay one has to change the
  signs of the three rate functions $ \hat{\Gamma}_U^P $, $ \hat{\Gamma}_L^P $
  and $ \hat{\Gamma}_F $. There are two more rate functions in the
  decay that were not discussed in this paper. They describe the relative 
  azimuthal angle dependence of the two decay planes that can be defined in the
  two-step decay process. Results on these additional azimuthal rate functions
  will be presented in a forthcoming publication.

  We have also provided a comprehensive discussion of the polarization of top
  quarks produced in $e^+e^-$-annihilation having the production of top quark
  pairs at future linear $ e^+e^- $-colliders in mind. As concerns polarized
  top decay, there are of course other areas of application of our decay 
  analysis. For example, polarized top quarks are also produced in single top
  quark production at hadron colliders \cite{ce}. Further, there is a high
  degree of correlation between the polarization of top and anti--top quarks
  produced in pairs either at $ e^+e^- $- \cite{cf,cg,ch} or hadron colliders
  \cite{ci} which can be probed through the joint decay distributions of the
  top and the anti-top.   
 \vspace{0.4cm}
 
  \noindent
 {\bf Acknowledgements:} M.~Fischer and M.C.~Mauser were supported by the
  DFG (Germany) through the Graduiertenkolleg "`Teilchenphysik bei hohen
  und mitt\-leren Energien"' at the University of Mainz. S.~Groote and
  J.G.~K"orner acknowledge partial support by the BMBF (Germany) under 
  contract 06MZ865.

\newpage


\vspace{1cm}
\centerline{\Large\bf Figure Captions}
\vspace{.5cm}
\newcounter{fig}

\begin{list}{\bf\rm Fig.\ \arabic{fig}:}{\usecounter{fig}
\labelwidth1.6cm\leftmargin2.5cm\labelsep.4cm\itemsep0ex plus.2ex}
\item Examples of polarization tuning of the top quark at $e^+e^-$-
      colliders through changes of the beam polarization with $ h^+ \!=\! 0 $
      and $ h^- $ ranging from $-1$ to $+1$ in steps of $ \Delta h^- \!=\! 0.2 $.
\item Normal component $ P^N $ of top polarization at $ O(\alpha_s) $. Shown are
      polarization values for $ h^+ \!=\! 0 $ ranging from $ h^- \!=\! -1 $
      to $ h^- \!=\! 1 $ in steps of $ \Delta h^- \!=\! 1 $.
\item Definition of the polar angles $ \theta_P $ and $ \theta $ in the top
      and in the $ W^+ $ rest frame, respectively.
\end{list}
\newpage


    
\thispagestyle{empty}
\begin{picture}(138,170)
  \put(-8,90){\setlength{\unitlength}{1mm}
  \begin{picture}(138,80)
    \centering \leavevmode 
    \psfig{file=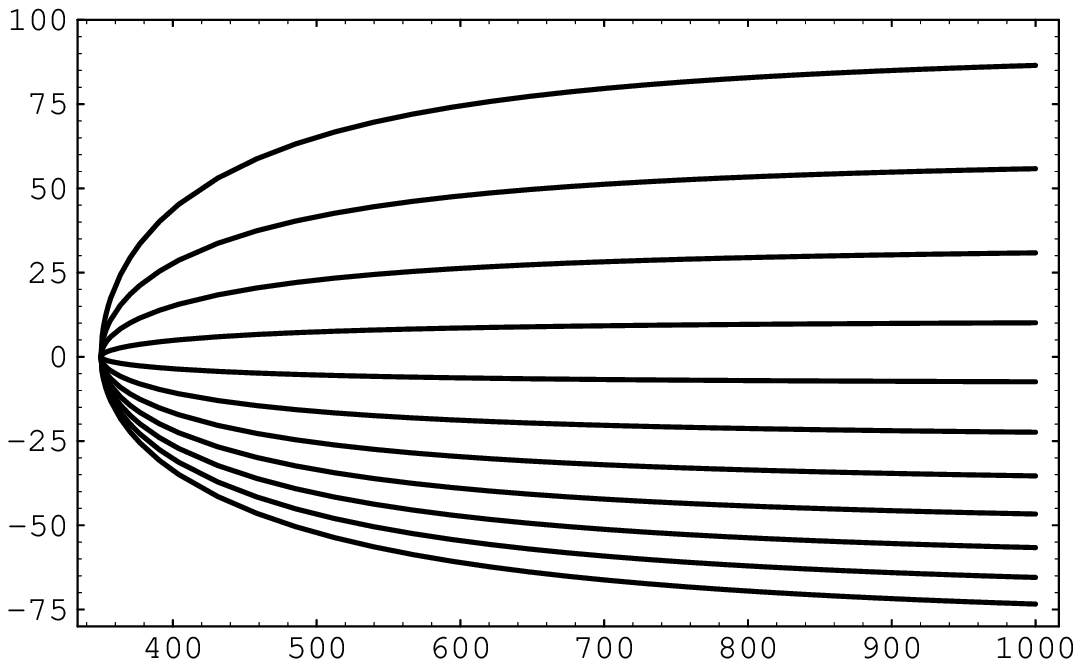,clip=} 
  \end{picture}}
  \put(72.0,95.0){\makebox(0,0){
  \textbf{center of mass energy $ \sqrt{q^2} $ [GeV]}}}
  \put(25.0,162){\makebox(0,0){
  \textbf{$ h^+ \!=\! 0 $ }}}
  \put(2,110.0){\rotateleft{\makebox{
  \textbf{long. polarization $ P^L $ [\%] }}}}
  \put(119,159){\makebox{$ \Ts{ h^- \!\!=\!+1.0 } $ }}
  \put(119,122){\makebox{$ \Ts{ h^- \!\!=\! \phantom{+} 0.0 } $ }}
  \put(119,104){\makebox{$ \Ts{ h^- \!\!=\!-1.0 } $ }}
  \put(-8,0){\setlength{\unitlength}{1mm}
  \begin{picture}(138,80)
    \centering \leavevmode 
    \psfig{file=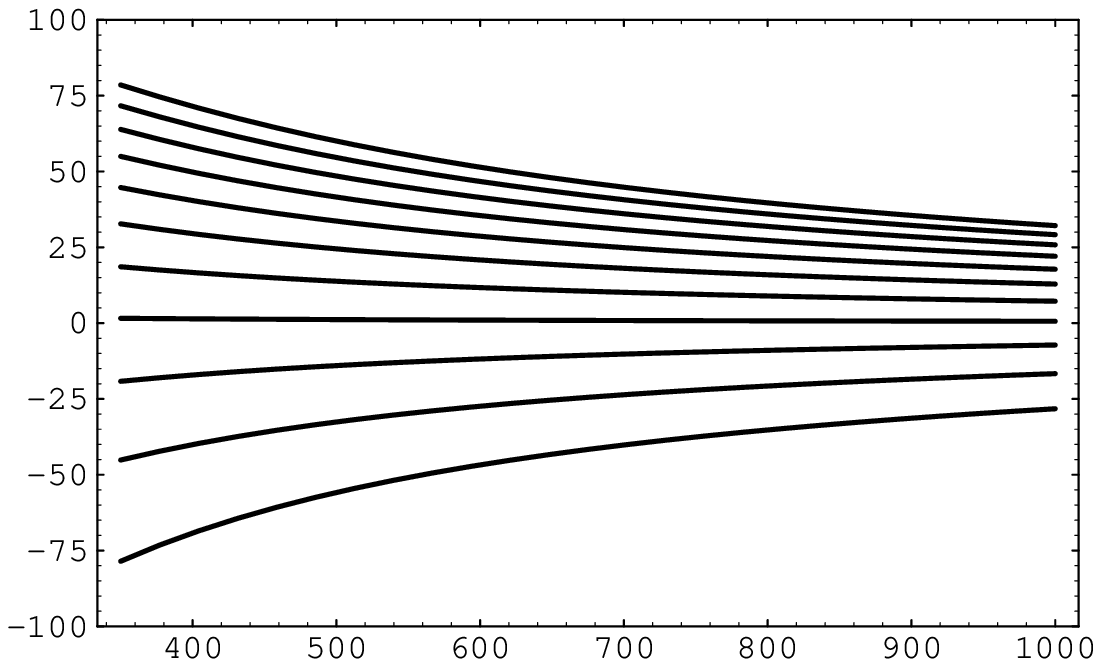,clip=}
  \end{picture}}
  \put(72.0,5.0){\makebox(0,0){
  \textbf{center of mass energy $ \sqrt{q^2} $ [GeV]}}}
  \put(25.0,072){\makebox(0,0){
  \textbf{$ h^+ \!=\! 0 $ }}}
  \put(2,20.0){\rotateleft{\makebox{
  \textbf{trans. polarization $ P^{\perp} $ [\%] }}}}
  \put(119,54){\makebox{$ \Ts{ h^- \!=\!-1.0 } $ }}
  \put(119,47){\makebox{$ \Ts{ h^- \!=\! \phantom{+} 0.0 } $ }}
  \put(119,33){\makebox{$ \Ts{ h^- \!=\!+1.0 } $ }} 
\end{picture}

\vspace{1truecm}
\centerline{\Large\bf Figure 1}
\newpage


\thispagestyle{empty}
\strut\vspace{5truecm}

\begin{picture}(138,90)
  \put(0,0){\setlength{\unitlength}{1mm}
  \begin{picture}(-20,0)
    \centering \leavevmode 
    \psfig{file=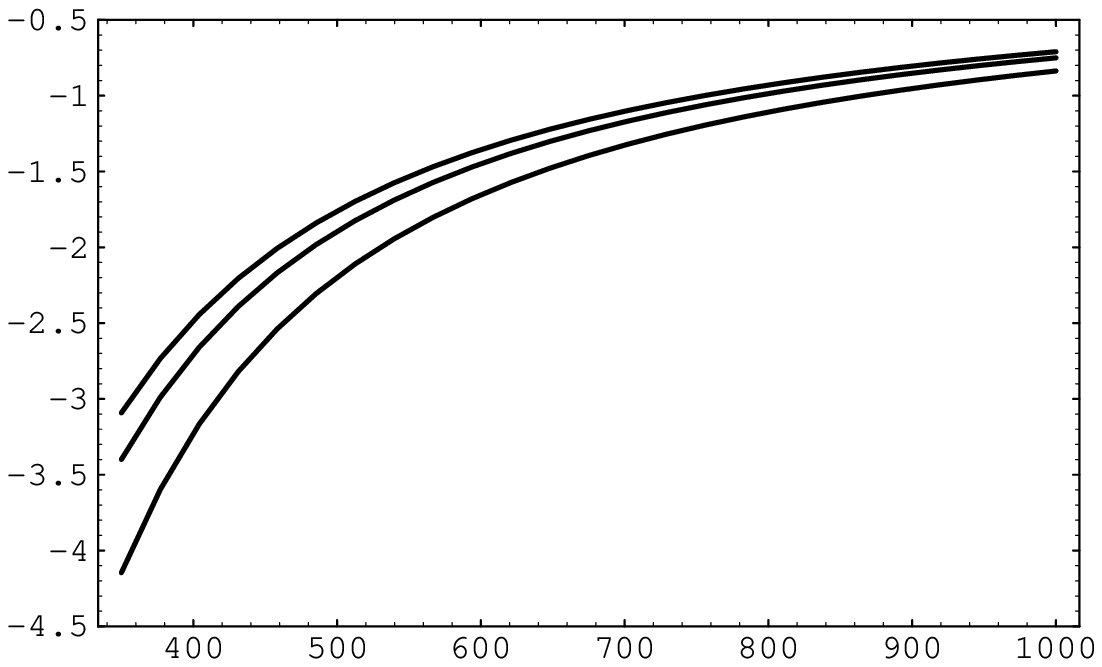,clip=} 
  \end{picture}}
  \put(80.0,05.0){\makebox(0,0){
  \textbf{center of mass energy $ \sqrt{q^2} $ [GeV]}}}
  \put(10.0,18.0){\rotateleft{\makebox{
  \textbf{normal polarization $ P^N $ [\%] }}}}
  \put(30.0,55.0){\makebox{$ \Ts{ h^- \!\!=\!-1.0 } $ }}
  \put(45.0,40.0){\makebox{$ \Ts{ h^- \!\!=\!+1.0 } $ }}
  \put(27.5,70.5){\makebox{$ h^+ \!=\! 0 $ }}
\end{picture}

\vspace{1truecm}
\centerline{\Large\bf Figure 2}
\newpage


\thispagestyle{empty}
\strut\vspace{6truecm}
\begin{figure}[h]
  \centering \leavevmode 
  \psfig{file=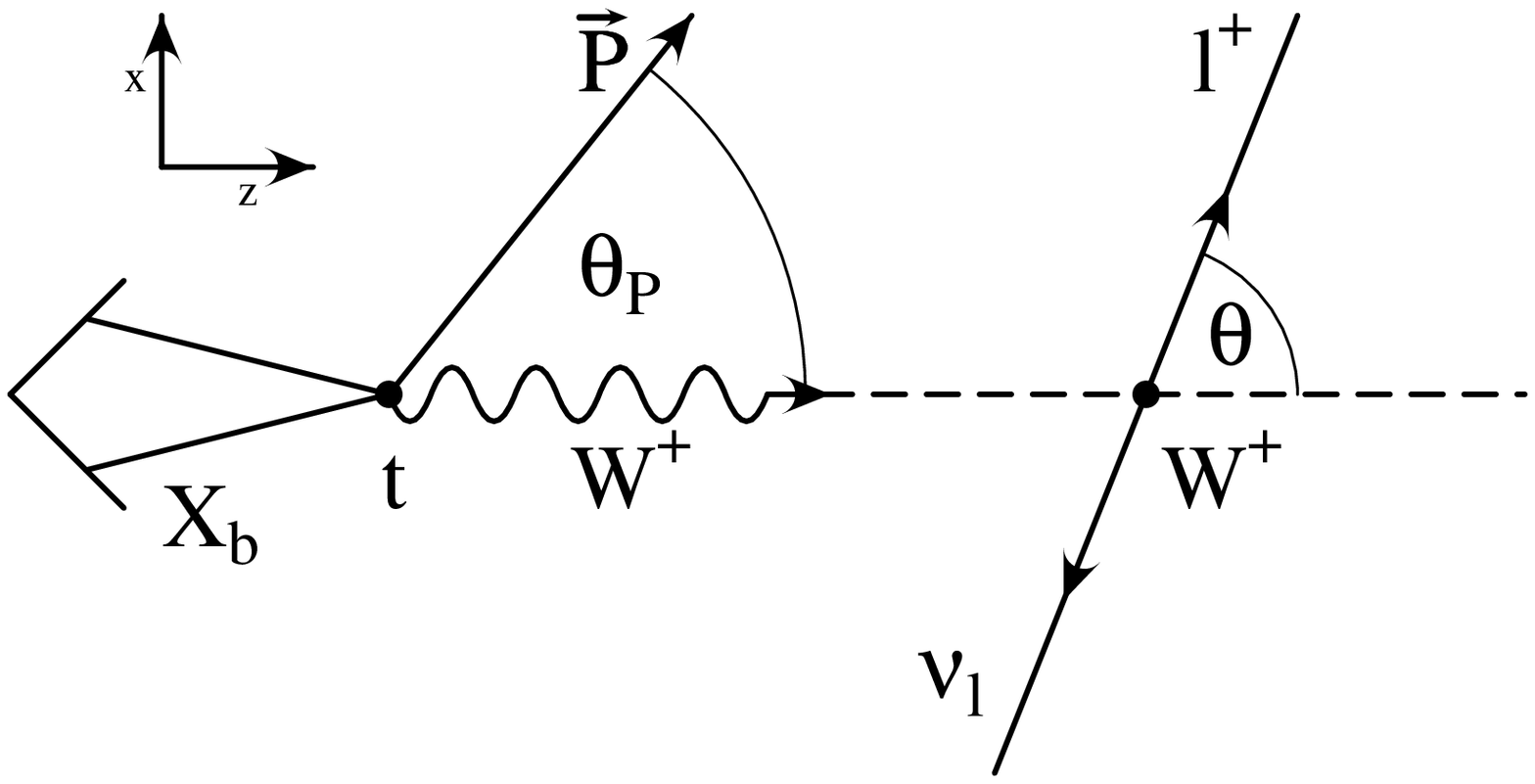,width=13cm,clip=}
\end{figure}

\vspace{1truecm}
\centerline{\Large\bf Figure 3}
\end{document}